\documentclass{article}
\usepackage{graphicx}

\begin{document}

\begin{center}
{\large\bf Concerning Measurement of Gravitomagnetism\\ in
Electromagnetic Systems\\}

\vspace*{5mm}

B.J. Ahmedov$^{a,b,c,d}$ and N.I. Rakhmatov$^{a,b}$\\

\vspace*{5mm}

{\it $^a$Institute of Nuclear Physics,
 Ulughbek, Tashkent 702132, Uzbekistan\\
$^b$The Abdus Salam International Centre for Theoretical Physics,
34014 Trieste, Italy\\ $^c$Inter-University Centre for Astronomy
and Astrophysics,\\ Post Bag 4, Ganeshkhind, Pune 411007, India\\
$^d$Ulugh Beg Astronomical Institute, Astronomicheskaya 33,
Tashkent 700052, Uzbekistan}
\end{center}
 \vspace*{5mm}

\begin{abstract}
{Measurement of gravitomagnetic field is of fundamental importance
as a test of general relativity. Here we present a new theoretical
project for performing such a measurement based on detection of
the electric field arising from the interplay between the
gravitomagnetic and magnetic fields in the stationary
axial-symmetric gravitational field of a slowly rotating massive
body.  Finally it is shown that precise magnetometers based on
superconducting quantum interferometers could not be designed for
measurement of the gravitomagnetically induced magnetic field in
the cavity of a charged capacitor since they measure the
circulation of a vector potential of electromagnetic field, i.e.,
an invariant  quantity including the sum of electric and magnetic
fields, and the general-relativistic magnetic part will be totally
cancelled by the electric one which is in good agreement with the
experimental results. }
\end{abstract}


\newpage

\section{Introduction}

General relativity predicts that a spinning central massive body creates
a gravitomagnetic field in addition to the Newtonian-like
monopolar gravitoelectric one, that is static gravitational mass
generates solely gravitoelectric field and a moving body, an
additional gravitomagnetic one as in electrodynamics (see, for
review, ~\cite{ignaz}). Almost all the classical tests of
general relativity (gravitational redshift, perihelion
precession of Mercury, gravitational deflection of light)
confirm the validity of the general-relativistic corrections to
the Newtonian gravitoelectric field with high accuracy.  But due
to the weakness of the gravitomagnetic field in the Solar system
it is very hard to detect it although many theoretical projects
on measuring gravitomagnetism have been proposed ~\cite{ignaz}.

In our recent papers~\cite{99pla,00adp} the gravitomagnetic effect on the
electric current and magnetic field has been investigated on the
theoretical level. In fact, the
influence of the angular momentum of the rotating gravitational source
may appear as a galvanogravitomagnetic effect in the current carrying
conductors~\cite{99pla} and as a general-relativistic effect of charge
redistribution inside conductors in an applied magnetic field
~\cite{00adp}.
It is natural to ask whether the gravitomagnetic interaction with
electric field can lead to the applicable general-relativistic effects.
Our aim here is to investigate an answer to this question and
present a new measurement scheme which could lead to a detection
of the earth's gravitomagnetic field in the electromagnetic
systems.

The paper is organized as follows: in section~\ref{em-capacitor}
we deal with electromagnetic fields in both the cavity of charged
capacitor and solenoid embedded in a space-time of slow rotating
gravitational object.  The next section~\ref{feasibility} is
devoted to a proposal for measuring gravitomagnetic field in the
electromagnetic systems.  In section~\ref{vasil} we discuss the
impossibility of detection of gravitomagnetic field through direct
measurement of a gravitomagnetically induced magnetic field in the
cavity of charged capacitor. Finally section~\ref{conclusion}
contains concluding remarks.

\section{Electromagnetic fields in electromagnetic systems in
space-time of rotating object}
\label{em-capacitor}

Consider first electromagnetic fields inside the charged capacitor in the
stationary axial-symmetric metric of a slowly rotating massive body as the
earth. Space-time outside a spherically symmetric mass $M$ with
the specific angular momentum $a=J/(cM)$ is described by the
Kerr metric~\cite{kerr}.  This differs from the Schwarzschild
solution for a static body by having non-diagonal terms, which
imply a local inertial frame to be rotating with respect to the
distant stars at infinity with the Lense-Thirring angular
velocity $\omega (r)\equiv {2aGM}/{c^2r^3}$~\cite{bahram}.  Then
the exterior metric of the slow rotating gravitational object
with mass $M$ (in the linear angular momentum $a$ approximation)
is
\begin{equation} ds^2=-N^2 c^2dt^2+N^{-2}dr^2+r^2d\theta
^2+r^2\sin {}^2\theta d\varphi ^2- 2{{\omega}} r^2\sin
{}^2\theta dtd\varphi \ , \label{eq:corot} \end{equation}
where $N\equiv\left({1-{2GM}/{\left(c^2r\right)}}\right)^{1/2}$, $G$
is the Newtonian gravitational constant. Thus the angular
velocity of the body as measured from the local free falling
(inertial) observer is ${\bar{\omega}} \equiv \Omega -\omega
(r)$, $\Omega$ is angular velocity of rotation of gravitational
object with respect to the distant stars.

Define the electric and magnetic
fields relative to an observer with four-velocity
$e^\alpha\equiv {dx^\alpha}/{d\tau}$ ($d\tau$ is its proper
time) as
\begin{equation}
\label{fields}
E_\alpha=F_{\alpha\beta}e^\beta\ ,\qquad
B_\alpha=-\frac{1}{2}
e_{\alpha\beta\mu\nu}e^\beta F^{\mu\nu} \ ,
\end{equation}
where
$e_{\alpha\beta\mu\nu}$ is the antisymmetric pseudo-tensorial
expression of the Levi-Civita symbol
$\epsilon_{\alpha\beta\mu\nu}$, $F_{\alpha\beta}=A_{\beta
,\alpha}-A_{\alpha ,\beta}$ and $A_\alpha$ are the tensor and
four-potential of electromagnetic field, respectively. Greek
indices run through $0,1,2,3$ and Latin indices from $1$ to $3$.

Element of an arbitrary 2-surface $dS^{\alpha\beta}$ can be represented in
the form
\begin{equation}
\label{surface}
dS^{\alpha\beta}= - e^\alpha\wedge m^\beta (ek)dS +
e^{\alpha\beta\mu\nu}e_\mu n_\nu\sqrt{1+(ek)^2}dS\ ,
\end{equation}
and the following couples
\begin{eqnarray}
m_\alpha =\frac{e_{\lambda\alpha\mu\nu}e^\lambda n^\mu k^\nu}
{\sqrt{1+(ek)^2}}, n_\alpha =\frac{e_{\lambda\alpha\mu\nu}
e^\lambda k^\mu m^\nu}{\sqrt{1+(ek)^2}}, k^\alpha=-(ek)e^\alpha+
\sqrt{1+(ek)^2}e^{\mu\alpha\rho\nu}e_\mu m_\rho n_\nu \nonumber
\end{eqnarray}
are established between the triple
$\{{\mathbf k},{\mathbf m},{\mathbf n}\}$ of vectors, $n^\alpha$
is normal to 2-surface, space-like vector $m^\alpha$ belongs to
the given 2-surface and is orthogonal to the four-velocity of
observer, a unit spacelike four-vector $k^\alpha$ belongs to the
surface and is orthogonal to $m^\alpha$, $dS$ is invariant
element of surface.

At the two-dimentional surface an arbitrary pair of mutually orthogonal
vectors $\{{\mathbf t},{\mathbf s}\}$ can be expressed through pair
$\{{\mathbf k}, {\mathbf m}\}$ in the following way
\begin{eqnarray}
          t^\alpha=\cos \varphi k^\alpha+\sin \varphi m^\alpha,\qquad
s^\alpha=\cos \varphi m^\alpha+\sin \varphi k^\alpha\ .\nonumber
\end{eqnarray}

Let a charged capacitor with the opposite surface density
of charges $\sigma_s$ at the plates be at rest in space-time
(\ref{eq:corot}); a constant charge is supplied by a battery of
constant electromotive force. Then the boundary conditions for
the jumps $[X]$ of the electromagnetic fields $E^\alpha,
B^\alpha$ and inductions $D^\alpha, H^\alpha$ at the plates
being the pieces of the coordinate surfaces take form:
\begin{eqnarray}
\label{bound}
&(n1)&\qquad [Bn]=0,\qquad
(n2)\qquad [Dn] = 4\pi\sigma_s\sqrt{1+(ek)^2}\ ,
\nonumber\\
&(t1)&\qquad [Et]=0,\qquad
(t2)\qquad [Ht] = 4\pi\sigma_s (ek) \sin\varphi\ .
\end{eqnarray}

The electromagnetic fields could be written in an orthonormal
frame with the tetrad $\{{\bf e}_{\hat \mu}\} = ({\bf e}_{\hat
0}, {\bf e}_{\hat r}, {\bf e}_{\hat \theta}, {\bf e}_{\hat
\phi})$
\begin{eqnarray}
\label{zamo_tetrad_0}
&&{\bf e}_{\hat
0}^{\alpha}  = N^{-1}\bigg(1,0,0,{\omega}\bigg) \ ,     \\
\label{zamo_tetrad_1}
&&{\bf e}_{\hat r}^{\alpha}  =
    N^{}\bigg(0,1,0,0\bigg) \ ,     \\
\label{zamo_tetrad_2}
&&{\bf e}_{\hat \theta}^{\alpha}  =
    \frac{1}{r}\bigg(0,0,1,0\bigg)  \ ,         \\
\label{zamo_tetrad_3}
&&{\bf e}_{\hat \phi}^{\alpha}  =
    \frac{1}{r\sin\theta}\bigg(0,0,0,1\bigg) \ .
\end{eqnarray}
The 1-forms $\{{\bf \omega}^{\hat
\mu}\} = ({\bf \omega}^{\hat 0}, {\bf
\omega}^{\hat r}, {\bf \omega}^{\hat \theta},
{\bf \omega}^{\hat \phi})$, corresponding to this
tetrad are
\begin{eqnarray}
\label{zamo_tetrad_1-forms_0}
&&{\bf\omega}^{\hat 0}_{\alpha}  =
    N^{}\bigg(1,0,0,0\bigg)\ ,          \\
\label{zamo_tetrad_1-forms_1}
&&{\bf\omega}^{\hat r}_{\alpha}  =
    N^{-1}\bigg(0,1,0,0\bigg)\ ,        \\
\label{zamo_tetrad_1-forms_2}
&&{\bf\omega}^{\hat \theta}_{\alpha}  =
    {r}\bigg(0,0,1,0\bigg)\ ,           \\
\label{zamo_tetrad_1-forms_3}
&&{\bf\omega}^{\hat \phi}_{\alpha}  =
    {r\sin\theta}\bigg(-{\omega},0,0,1\bigg)\ .
\end{eqnarray}

Suppose that the plates of capacitor are pieces of the
coordinate surfaces $r = const.$ with the characteristic
vectors
\begin{eqnarray}
k^\alpha \{0,0,0,\frac{1}{r\sin \theta}\}\ ,
\quad m^\alpha  \{0,0,-\frac{1}{r},0\}\ ,
\quad n^\alpha \{0,{N},0 ,0\}\ ,
\end{eqnarray}
where the boundary conditions (\ref{bound}) take the form
\begin{eqnarray}
 [B^{\hat r}] = [E^{\hat\theta}] = [E^{\hat\phi}]
= [H^{\hat\phi}] &=& 0 ,\nonumber\\
{[D^{\hat r}]}= 4\pi\sigma_s{N}^{-1}|_{r=const}, \qquad
[H^{\hat\theta}] &=& 4\pi\sigma_sN^{-1}\frac{\omega r\sin \theta}
{c} |_{r=const}\ .
\end{eqnarray}

Due to the axial symmetry and stationarity of the problem the field
vectors depend on the space coordinates $r$ and $\theta$.
If there is the potential difference $\Delta\phi\equiv\Delta A_0$ between
the plates of capacitor $r=r_1$ and $r=r_2$ ($\Delta r=r_2 -r_1$)
then nonvanishing components of the electromagnetic
field are defined by the formulae:
\begin{eqnarray}
\label{field}
E^{\hat r}=\frac{\Delta\phi}{\Delta r}N^{-1},\quad
B^{\hat\theta}=\frac{\omega r\sin\theta}{c}
\frac{\Delta\phi}{\Delta r}N^{-1} \ .
\end{eqnarray}

As one can see the electric field in the metric~(\ref{eq:corot}) is
modified by the coupling to the static monopolar part of
gravitational field defined by the mass $M$ while the magnetic
field is generated due to the interplay between the dragging of
inertial frames and electric field, and vanishes in the
Schwarzschild space-time.

Consider now electromagnetic fields outside infinitely long
solenoid carrying constant electric current. It has been
recently shown (see, for example,~\cite{muslimov,ram})
that the electric field being proportional to $(\omega r/c)B$ is
induced in a slowly rotating metric (\ref{eq:corot})  of a
star with magnetic field $B$. Analogously one can expect that
in the linear approximation in the angular velocity of rotation
the electric field
\begin{equation}
\label{EF}
E\approx (\omega r/c)B
\end{equation}
is induced around the solenoid carrying constant
current.

\section{Feasibility of an experiment on measuring gravitomagnetism}
\label{feasibility}

It is well-known that the electric field induced by gravity
was first detected in~\cite{wf,wf71,lwf} by analyzing the
time-of-flight distribution of electrons falling freely within a
metal cylindrical hollow tube (with the length $L$) where the
electrons experienced the force of gravity $mg$ ($g$ is the
gravitational acceleration), the gravitoelectrically induced
electric field $E_i$ and an applied weak uniform electric field
$E_a$. By measuring the maximum observable flight time of
electrons \begin{equation} t_{max}=[2Lm/(mg-eE_i+eE_a)]^{1/2}
\end{equation}
for several values of $E_a$, the experimental value
for $E_i$ was obtained with high accuracy up to
$10^{-11}-10^{-12}V/m$.

In the present paper a simple method for proposal on measuring
gravitomagnetism is studied by assuming ideal experimental
conditions.  In order to grasp the idea, let us start from the
equation of motion of charge $q$ with mass $m$ in
electromagnetic and stationary gravitational field of a rotating
mass (in the weak gravitational field and slow motion
limit)~\cite{ignaz}
\begin{equation}
m\frac{d{\mathbf v}}{dt}\cong q({\mathbf E}
    +\frac{1}{c}{\mathbf v} \times
    {\mathbf B}) + m({\mathbf E}_g+
    \frac{1}{c}{\mathbf v}\times{\mathbf B}_g)\ ,
\label{eq:gmot}
\end{equation}
where the gravitational field is decomposed into a ``gravitoelectric"
part \begin{equation} {\mathbf E}_g=-\frac{GM}{r^2}\hat{\mathbf
r} \end{equation} and a ``gravitomagnetic" one \begin{equation}
{\mathbf B}_g={\mathbf \nabla}\times{\mathbf A}_g=
\frac{2G}{c}[\frac{({\mathbf J}- 3({\mathbf J\cdot\hat
 r})\hat{\mathbf r})}{r^3}]\ , \label{eq:gm} \end{equation}
$\hat{\mathbf r}$ is the
unit vector responsible for the position of test particle and
the gravitomagnetic potential is given by
\begin{equation}
{\mathbf A}_g=-\frac{2G}{c}\frac{{\mathbf J}\times{\mathbf r}}{r^3}\ .
\end{equation}

In the frame $K\{x,y,z\}$ whose $(x,y)$-plane coincides with the
equatorial plane of gravitating source its proper angular
momentum is directed along the $z$- axis,
${\mathbf J}=J{\mathbf i}_z$ and the equation for the gravitomagnetic
field ${\mathbf B}_g$ immediately yields
\begin{equation} {\mathbf B}_g=
\frac{2G}{c}\frac{J}{R^3}{\mathbf i}_z\ .
\label{gm}
\end{equation}

The solution of the equation of motion~(\ref{eq:gmot}) for a charged
particle, when there is no gravity, in crossed constant electric and
magnetic fields is well-known (see, for example,~\cite{landau}). The crucial
point for further analysis is that in the direction perpendicular to
the common plane of fields $\mathbf E$ and $\mathbf B$, a particle moves
with a velocity which is a periodic function of time, drift velocity
\begin{equation}
{\mathbf v}_{drift}=\frac{{\mathbf E}\times{\mathbf B}}{B^2}\ .
\end{equation}

In the particular cases i) ${\mathbf B}=0, {\mathbf v}_{drift}=0$ in
the perpendicular direction (the motion only in the direction being
parallel to electric field); ii) ${\mathbf E}=0, {\mathbf v}_{drift}=0$
and the orbit of the particle is a helix, with its axis parallel to
magnetic field.

Thus according to
equations~(\ref{eq:gmot}) and (\ref{field}) the motion of electrons
in the vicinity of the solenoid carrying constant electric
current will be effected by the gravitoelectric field and
gravitomagnetically induced electric field. The investigation of
the electron motion in the vicinity of the solenoid can give a
successful method to detect the gravitomagnetic field of the
earth. The essential part of the suggested method could be based
on the gravitomagnetical generation of electric field crossed to
the original magnetic one.

Now let us have a look at figure. The magnetic field
$B^{\hat\alpha}\{0,0,B\}$ is along the
horizontal $z$ axis and gravitomagnetically induced electric
field (\ref{EF})
$E^{\hat\alpha}\{0,(\omega R/c)B,0\}$ is along the
vertical $y$ axis. The particle moves in the $z$ direction with a
uniform velocity. Then in the coordinate system $K(x,y,z)$, the equation
of motion of charged particle reads
\begin{equation}
m\frac{d{\mathbf v}}{dt}=m{\mathbf E}_g+
q{\mathbf E}+q{\mathbf v}
\times {\mathbf B}_t \ ,
\label{motion}
\end{equation}
with the gravitoelectric field, velocity and electromagnetic field
given by
\begin{eqnarray}
\label{polya}
{\mathbf E}_g&=& g{\mathbf j}\ ,\qquad
{\mathbf v}= \dot{x}{\mathbf i}+ \dot{y}{\mathbf j}+ \dot{z}{\mathbf k}\ ,
\nonumber\\
{\mathbf E}&=& E_x{\mathbf i}+ E_y{\mathbf j}+ E_z{\mathbf k}\ ,
\qquad {\mathbf B}= B_t{\mathbf k}\ ,
\qquad B_t = B+\frac{m}{q}B_g \ ,
\end{eqnarray}
where ${\mathbf i}, {\mathbf j}$ and ${\mathbf k}$ are unit
vectors along the corresponding rectangular axes (see Figure 1).

\begin{figure}[hbtp] \label{fig1}
\includegraphics[height=86mm, width=140mm]{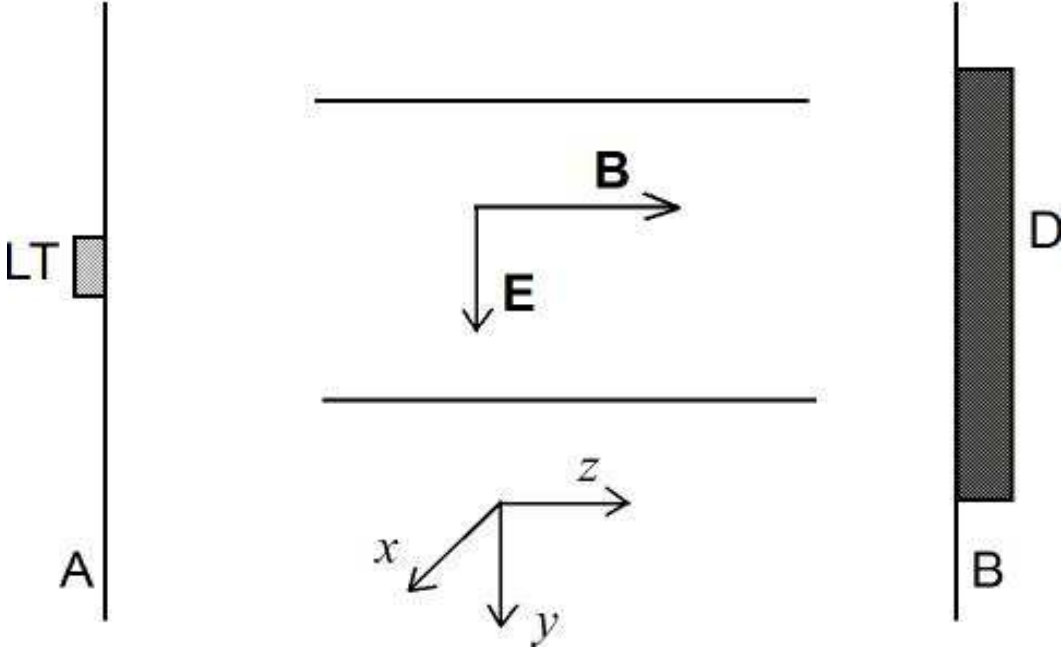}
\caption{Scheme of proposed experiment. $D$ is the detector,
${\mathbf B}$ is the magnetic field, ${\mathbf E}$ is the electric
field.}
\end{figure}

Using the equation of motion~(\ref{motion}) and~(\ref{polya}) a very
simple calculation leads to the system of equations
\begin{eqnarray}
\label{system}
&&\frac{d}{dt}\left(\dot{x}+i\dot{y}\right)+\frac{qB_t}{m}
\left(\dot{x}+i\dot{y}\right) = \frac{q}{m}
\left(E_x+iE_y\right)+ig\ ,
\\\nonumber
&&\ddot{z}=\frac{q}{m}E_z \ .
\end{eqnarray}

It is an ideal case to have $E_x=E_y=0$ in the system~(\ref{system}).
However there are always small stray electric fields (In a
superconducting drift tube as demonstrated by Witteborn and
Fairbank~\cite{wf}, stray electric fields could be controlled
at the level of $10^{-11}V/m$.).

In the real experimental situation a confining electric field can be
produced. We shall also include some small constant stray fields
$e^{\hat i}\{e_x,e_y,e_z\}$, so that for electric field in the
system~(\ref{system}) we have
\begin{equation}
\label{electric}
E_x=-c_1z-e_x\ ,\qquad
E_y=-c_2z-e_y\ ,\qquad
E_z=-c_3z-e_z\ .
\end{equation}

Introducing electric field (\ref{electric}) into (\ref{system})
gives
\begin{eqnarray}
\label{system1}
&&\frac{dv}{dt}+i\Omega_L v=-\frac{q}{m}\left(c_1+ic_2\right)z
-\frac{q}{m}\left(e_x+ie_y\right) +ig\ ,
\\\nonumber
\label{system2}
&&\ddot{z}+\Omega_0^2z=-\frac{q}{m}\
\end{eqnarray}
with
\begin{equation}
\Omega_L=\frac{qB_t}{m}\ ,\qquad
\Omega_0^2=\frac{qc_3}{m}\ ,\qquad
v=\left(\dot{x}+i\dot{y}\right)\ .
\end{equation}

The general solution of the equation (\ref{system2}) is
\begin{equation}
\label{sol_z}
z=C_1\sin\left(\Omega_0 t+\varphi\right)-\frac{e_z}{c_3}\
\end{equation}
and charge oscillates along $z$ axis.
Here $C_1$ and $\varphi$ are integration constants depending on
boundary conditions.
According to the solution~(\ref{sol_z}), the equation (\ref{system1})
can be written as
\begin{equation}
\label{eqn_v}
\frac{dv}{dt}+i\Omega_L v=-
A\sin\left(\Omega_0 t+\varphi\right)+B_t +ig\ ,
\end{equation}
with
\begin{equation}
A=C_1\frac{q}{m}\left(c_1+ic_2\right)\ ,
\qquad
B_t=\frac{q}{m}\left(-\left(e_x+ie_y\right)+
\frac{e_z}{c_3}\left(c_1+ic_2\right)\right)\ .
\end{equation}

The solution of equation (\ref{eqn_v}) is
\begin{equation}
\label{sol_v}
v=-\frac{A}{\Omega_L(1-\Omega_0^2/\Omega_L^2)}
\left(\frac{\Omega_0}{\Omega_L}\cos\left(\Omega_0 t+
\varphi\right)-i\sin\left(\Omega_0 t+\varphi\right)\right)
+\frac{a}{\Omega_L}+\frac{B_t}{i\Omega_L}+C_2e^{-i\Omega_Lt}\ .
\end{equation}

As a consequence of (\ref{sol_v}), $\dot{x}$ and $\dot{y}$ are the
periodic functions of time with average values
\begin{equation}
\label{sol_period}
\bar{\dot{x}}=\frac{a}{\Omega_L}+\frac{q}{m\Omega_L}
\left(\frac{e_z}{c_3}c_2-e_y\right)\quad
\bar{\dot{y}}=\frac{q}{m\Omega_L}
\left(\frac{e_z}{c_3}c_1-e_x\right)\ .
\end{equation}

In the case when stray electric fields $e^{\hat i}=0$ vanish,
the solution (\ref{sol_v}) becomes very simple:
\begin{equation}
\bar{\dot{x}}=\frac{a}{\Omega_L}
\qquad and \qquad
\bar{\dot{y}}=0\ .
\end{equation}

With a good control of stray fields as in paper~\cite{wf71}
drift along the $x$ axis caused by field (\ref{EF}) could be
measured.

\section{On direct measurement of gravitomagnetically induced
electromagnetic field}
\label{vasil}

It should be mentioned that
the question of direct measurement of gravitomagnetically
generated magnetic field has been experimentally tested~\cite{vas78}
and the null
result of the experiment was treated as a verification of the principle of
equivalence.

In experiment~\cite{vas78}  superconducting quantum interferometers
(SQUIDs)
have been used in order to detect the magnetic field generated by the
gravitomagnetic effect on electric field, i.e. to measure magnetic field
inside charged capacitor embedded in the earth's gravitational field
under the different space orientations of the equipment.
The original idea of experiment~\cite{vas78}  was based on the following
 two main assumptions.

First, the constitutive relations~\cite{landau}
\begin{eqnarray}
{\mathbf B}=\frac{\mu{\mathbf H}}{\sqrt{-g_{00}}}+({\mathbf   g}\times
{\mathbf   E}),   \qquad
{\mathbf   D}=\frac{\epsilon{\mathbf   E}}{\sqrt{-g_{00}}}-({\mathbf g}
\times{\mathbf  H})
\end{eqnarray}
between the electromagnetic fields and inductions
for isotropic media with dielectric
permittivity $\epsilon$ and the magnetic permeability $\mu$
lead to the assumption that under the influence of the dynamical
gravitomagnetic part
$g_\alpha\{0,{g_{0i}}/{\sqrt{-g_{00}}}\}$ of the gravitational field
on  electromagnetic phenomena a magnetic
field will be induced by the stationary electric field and vice versa.

Second, a SQUID was expected to measure the
flux of magnetic induction  $\int{\mathbf B} d{\mathbf S}$
in general relativistic case as in the flat space-time, that is the
standard electromagnetic result according to which a SQUID measures the
flux of a magnetic field was applied to the treatment in the space-time
of rotating gravitational object.

Assuming the earth in the first approximation is slowly rotating
gravitational body~\footnote{The authors of the paper~\cite{vas78}
assumed that the maximal value of $g_\alpha$ may reach $10^{-9}-
10^{-10}$ due to the rotation of the Solar system together with
the earth relative to the centre of galaxy whereas
$g_\alpha\approx 10^{-9}$ due to the diurnal rotation of the earth
as gravitational source.} and using these two prepositions the
authors~\cite{vas78} expected that SQUID should detect the
nonvanishing gravitomagnetically generated quantity $\int({\mathbf
g}\times{\mathbf E})d{\mathbf S} $ inside the charged capacitor
located on the earth. However, a number of measurements of the
magnetic flux inside the charged capacitor on the earth did not
give positive result. It was established experimentally that the
quantity measured by the SQUID inside the condenser does not
depend on the space orientation of the equipment with respect to
the distant stars.

In order to show that null experiment~\cite{vas78}
can be  predicted  in the framework of the general-relativistic
electrodynamics of continuous media it is necessary to get an
answer to two important questions:
(i) which quantity is measured by the SQUID embedded in external
gravitational field and
(ii) how the measuring quantity corresponds to the electromagnetic
    fields generated in the cavity of the charged capacitor.

     In order to get a solution to the first question we consider the
general-relativistic generalization of the Bohr-Sommerfeld quantization
condition for the Cooper pairs
\begin{eqnarray}
\label{bohr}
\oint_{\gamma}^{} p_\alpha dx^\alpha = \oint_{\gamma}^{} (2mcu_{(s)\alpha}
+\frac{2e}{c} A_\alpha)dx^\alpha = n\hbar\ ,
\end{eqnarray}
which is the macroscopic analog of the quantization of angular
momentum $p_\alpha$ in an atomic system. Here $n$ is an integer, $\hbar$
is the Planck's constant, $\gamma$ is a closed curve in the interior of
superconductor, $u_{(s)\alpha}$ is the
four-velocity of the Cooper pairs.

Using the Stokes theorem~\cite{mtw}  we can extend the integration over
the surface
$dS^{\alpha\beta}$ spanned by the curve. Then condition (\ref{bohr}) may be
rewritten in the form
\begin{eqnarray}
\frac{mc^2/e}{\sqrt{1-v_{(s)}^2/c^2}}\int A_{\alpha\beta}dS^{\alpha\beta}+
\Lambda^{-1}c\oint \hat\jmath_{(s)\alpha}dx^\alpha +\frac{1}{2}\int
F_{\alpha\beta}dS^{\alpha\beta}=n\Phi _0\ ,
\end{eqnarray}
which asserts, due to nonpenetration of the superconducting current
into the
bulk of the superconductor, that the sum of flux of the electromagnetic
tensor and the purely relativistic term arising
from the rate of rotation
$2A_{\beta\alpha}=u_{\alpha ,\beta}-u_{\beta ,\alpha}+u_{\beta}\wedge
w_{\alpha}$ is quantized in the general-relativistic context
(the general-relativistic contribution leads to the relativistic London
moment~\cite{dewitt}-\cite{anandan}). Here
$\Phi_0={\hbar c}/{2e}$ is the flux quantum, $\hat\jmath^\alpha_{(s)}$ is
the superconducting current,  $v^\alpha_{(s)}$ is the velocity of
superconducting electrons relative to the medium as a whole with the
four-velocity $u^{\alpha}$ and the absolute acceleration
$w_\alpha =u_{\alpha ;\beta}u^\beta$, $\wedge$ is the symbol of exterior
product.

Thus the superconducting quantum interferometers are adjusted for
measurement of circulation of the 4-potential
of electromagnetic field along the closed contour
\begin{eqnarray}
\label{flux}
\Phi_B=\oint A_\alpha dx^\alpha = \frac{1}{2}\int F_{\alpha\beta}
dS^{\alpha\beta}\ ,
\end{eqnarray}
and consequently the SQUID in principle measures the flux of
electromagnetic field tensor $F_{\alpha\beta}$ (rather the
flux of magnetic field as expected in~\cite{vas78}) which is defined
in an invariant way and
does not explicitly depend on an observer.

Magnetic flux (\ref{flux}) through the surface of a measuring contour
may be nonvanishing only if it contains the projection on
the surface $\theta = const$ with the characteristic vectors
\begin{eqnarray}
k^\alpha \{0,0,0,\frac{1}{r\sin\theta}\},\quad
 m^\alpha \{0,N,0 ,0\},\quad
n^\alpha \{0,0,\frac{1}{r},0\}\ .
\end{eqnarray}
Then having in mind (\ref{fields}), (\ref{field}) and (\ref{surface})  we
obtain
\begin{eqnarray}
\label{zero}
\Phi_B=\int B_\alpha n^\alpha\sqrt{1+(ek)^2}dS+\int E_\alpha m^\alpha
(ek)dS=\nonumber\\
\frac{1}{c}\int (\Delta\phi/\Delta r) N^{-1}\omega
\sin \theta dS -\frac{1}{c}\int (\Delta\phi/\Delta r) N^{-1}
\omega\sin \theta dS \equiv 0\ .
\end{eqnarray}

Similarly one may show that the value of the flux of electromagnetic
tensor (\ref{flux}) through the surface of measuring loop inside the
charged capacitor is equal to zero for any other space orientations of
equipment in the axisymmetric space-time geometry~(\ref{eq:corot}).
Thus the total flux inside the capacitor being
independent of its orientation is in the agreement with the null result
of the treated experiment~\cite{vas78} (unlike Vasil'ev's [12]
treatmeant of the null
result as a consequence of absence of the gravitatomagnetic effects on
the electromagnetic processes according to the principle of
equivalence) which indirectly confirms that the magnetic field can be
created inside the capacitor by the interplay between electric and
gravitomagnetic fields. It appears the SQUID measures the invariant
flux (\ref{zero}) including contribution from the electric field and
consequently the contribution due to the general-relativistic
gravitomagnetic effect, i.e., the second term on the
right hand side of (\ref{zero}) is totally cancelled by the first term
in the process of the measurement.
It should be noted the obtained result could be considered as a
consequence of (\ref{flux}) when the four-potential
$A_\alpha\{A_0,0,0,0\}$ of electromagnetic
field has only nonvanishing component $A_0\ne 0$.

Consider now a possibility to detect the gravitomagnetic field in the
space of the solenoid with constant electric current. It is known
that the constant magnetic field in a slowly rotating metric
(\ref{eq:corot})  is modified by the gravitoelectric factor but
has no any contribution from the gravitomagnetic field
(see, for example,~\cite{muslimov,ram}). In the linear
approximation in the angular velocity of rotation magnetic flux
(\ref{flux}) around the solenoid carrying constant current is
defined only by the magnetic field since the term produced by
the electric field in (\ref{flux}) is proportional to
$O(\omega^2)$ and therefore negligible. So the gravitomagnetic
field can not be detected via the direct measurement of the
elecromagnetic field around solenoid by superconducting loop
although it can be tested via the general-relativistic effect of charge
redistribution inside a conductor embedded in the magnetic field of the
solenoid~\cite{00adp}.
Thus we can make a general conclusion that the direct SQUID's measurements
of the electromagnetic fields in the stationary conditions can not detect
the gravitomagnetic field although in our recent paper~\cite{99pla}
we have shown that the SQUID measuring the galvanogravitomagnetic current
(not field) can be used as a detector of the gravitomagnetism.

\section{Concluding remarks}
\label{conclusion}

As we have shown here, on one side, due to the gravitomagnetic effect,
a magnetic field~
(\ref{field}) will be induced inside a cavity of condenser in addition to
the electric one. On other side, the second term arising on the right hand
side of the flux~(\ref{zero}) totally compensates the gravitomagnetically
induced magnetic field. In this sense the experiment~\cite{vas78}
is analogous to the null result experiment~\cite{jain} on test of
the principle of equivalence for electromagnetic systems, where the
gravitational redshift of the microwave frequency $\Delta\nu/\nu=\lambda$
was totally compensated by the general-relativistic effect due to the fact
that the gravitoelectrochemical potential
${\tilde{\mu_e}}=\mu_e(1+\lambda )$
(rather than the electrochemical one $\mu_e$) is constant during
thermodynamical equilibrium (parameter $\lambda =Lg/c^2$ depends on the
height difference $L$ and the acceleration due to gravity $g$).
However both compensating general-relativistic effects were separately
detected in a number of experiments, for example, the gravitational
redshift for the frequency by Pound and Rebka~\cite{pound} and
the electric field inside conductors produced by the inhomogeneity of the
chemical potential in the gravitational field by Witteborn and
Fairbank~\cite{wf,wf71,lwf}.

The advantage of the time-of-motion experiment over the Vasil'ev's one
~\cite{vas78} is that a non-null result is predicted by general relativity
as opposed to the null result for the latter experiment. Also in the
proposed experiment, with the time-of-motion technique, Newtonian physics
would predict a null result, whereas in the proposed experiment it would be
necessary to assume the existence of the earth's gravitomagnetic field to
distinguish between the results predicted by general relativity and
Newtonian gravity.

It should be mentioned that on earth, the angular velocity
of the capacitor with respect to a local inertial frame
${\mathbf\Omega}_{L}$ is
given by~\cite{ignaz}:

$${\mathbf\Omega}_{L}={\mathbf\Omega}_{VLBI}-{\mathbf\Omega}_{Th}-
{\mathbf\Omega}_{S}-{\mathbf\Omega}_{LT},$$

where ${\mathbf\Omega}_{VLBI}$ is the angular velocity of the laboratory
with
respect to an asymptotic inertial frame, ${\mathbf\Omega}_{Th}$ and
${\mathbf\Omega}_{S}$ are, respectively,
the contributions of the Thomas precession arising from
non-gravitational
forces and of the de Sitter or geodetic precession. As a result, in
order to detect
${\mathbf\Omega}_{LT}$ (which is $\omega$ in metric~(\ref{eq:corot}))
one should measure ${\mathbf\Omega}_{L}$ and then
substract from it the independently measured value of
${\mathbf\Omega}_{VLBI}$
with VLBI (Very Long Baseline Interferometry, see, for example,
~\cite{koval}) and the contributions
due to the Thomas and de Sitter precessions.

The typical parameters for earth which can be considered as uniformly
rotating body are
$a_\oplus=({2}/{5}) \Omega_\oplus R_\oplus^2$, $2\times
G{M_\oplus}/{c^2}=2\times 0.44cm$, $R_\oplus\approx 6.37\times
10^8cm$, $\Omega_\oplus=7.27\times 10^{-5}rad/s$ and consequently
$\Omega_{LT}\approx 10^{-9}\Omega_\oplus$ is extremely weak.
The experiment is quite difficult since the measured effect is small
compared to potential disturbances and careful attention must be paid to
possible systematic experimental errors coming from seismic accelerations,
local gravitational noise, earth's electromagnetic field,
atmospheric disturbances etc.
We mention only how  few major difficulties can be reduced: (i)
shielding the earth's electric field with help of Faraday cage with
high accuracy, (ii) shielding the earth's
magnetic field by means of superconducting shells, and (iii) vacuum
measurements to prevent the effects arising from the air ionization.

We do not provide here any numerical estimates for the experimental
technique since we would like only to underline the existence
of the possibility of detecting
the gravitomagnetically generated electric field. The detailed
design and study of the method under real experimental
conditions is under our consideration. However for
some technical details and study of unwanted effects as
electrostatic effects, magnetostatic effects, the patch effect,
the electron and the lattice sag effects, thermal and vacuum
effects one can look to~\cite{wf71}.

From our research, we can conclude that (i) the effect of the
gravitomagnetic force on the elecrostatic (magnetic) field is to
induce a magnetic (electric) field, (ii) gravitomagnetically
generated magnetic field can not be directly measured by the SQUID
which is in agreement with the results of the null
experiment~\cite{vas78}, (iii) the electric field arising from the
interplay between the general relativistic gravitomagnetic and
magnetic fields seems to be experimentally verifiable with the
measurement of the trajectory-of-motion of charged particles (see
equations~(\ref{sol_period})).

\section*{Acknowledgements}

B.A. greatly acknowledges the hospitality at the IUCAA, Pune and
the Third World Academy of Sciences for financial support towards
his visit to India in the spring 2002. The research is supported
by the grants AC-83 and NET-53 through the Office of External
Activities of AS-ICTP. This research is also supported in part by
the UzFFR (project 01-06) and projects F.2.1.09, F2.2.06 and
A13-226 of the UzCST.

\end{document}